# Introducing High School Students to Version Control, Continuous Integration, and Quality Assurance

Joseph Latessa
Department of Computer Science
Wayne State University
Detroit MI USA
jlatessa@wayne.edu

Aadi Huria
Senior
Salem High School
Canton MI USA
huria.aadi@gmail.com

Deepak Raju
Senior
Salem High School
Canton MI USA
Deepak.Raju294@outlook.com

## ABSTRACT

Software Engineering concepts such as version control, continuous integration, and unit testing are often not presented in college computer science curriculums until the third year of study, after completing several semesters of programming courses. Throughout the summer of 2023, two high school students volunteered in our lab at Wayne State University where I'm a graduate research assistant and Ph.D. student in computer science. The students had taken AP Computer Science but had no prior experience with software engineering or software testing. This paper documents our experience devising a group project to teach the requisite software engineering skills to implement automated tests that meaningfully contribute to open-source scientific computing projects developed in connection with our lab. We describe the concepts covered, tools used, and software tests written in this early introduction to software engineering while maintaining shared emphases on education and the deployment of our work.

## 1 INTRODUCTION

During the summer of 2023, two high school students volunteered to assist in one of Wayne State University's computer science research labs, which focuses on high-performance computing and software engineering for scientific applications. The students aspire to study computer science as their major when they begin their undergraduate studies in 2024. Prior to volunteering in our lab, they had successfully completed AP Computer Science courses where they learned Java and Python but had no prior experience with software engineering concepts such as version control, continuous integration, and unit testing. Our goals were to develop a summer project that would be appropriate for the students' current experience and education, provide a learning opportunity to gain new skills relevant to assisting in a laboratory environment, and to provide an opportunity to produce work that meaningfully contributes to the scientific computing applications developed in connection with our lab.

Two scientific computing applications developed in connection with our lab include JETSCAPE [1] and GOMC [2]. JETSCAPE is a multi-institutional collaboration that develops event generators to simulate ultra-relativistic heavy-ion collisions. GOMC is an application to simulate molecular systems using the Metropolis Monte Carlo Algorithm [3]. Both projects are open-source and maintain their respective repositories on GitHub. They also use GitHub Pages to maintain and deploy their project websites.

To meet our project goals, we decided to emphasize automated software testing. JETSCAPE [1] and GOMC [2] have previously developed and deployed some tests, but there was an interest among the collaborators to further expand upon the tests in place. Since the student volunteers had not yet studied software engineering and were not yet familiar with concepts such as continuous integration and unit testing, attaining these skills while working on this project would provide a unique educational value. Since the students had already completed AP Computer Science and were already comfortable with the Java and Python languages, the students could apply the programming skills they already learned to develop these automated tests.

Although having studied Java and Python, the students had no prior experience with Linux or with working at the command line. They also had no previous experience working with Git or any version control platform. This project would therefore have to begin with preliminary introductions to the Linux shell environment and version control. Since continuous integration would also be a new concept, a demonstration of GitHub Actions and writing YAML files would conclude our preliminary introduction.

The main substantive tasks of the project would be the implementation of three automated tests. The first two tests, an HTML validator, and a link checker, would be deployed on the JETSCAPE [1] and GOMC [2] websites. The implementation of the third test would require discussing the concept of unit testing and then deploying a series of unit tests to ensure the correctness of JETSCAPE's XML formatted input.

The tools used throughout the project include Visual Studio Code, the Ubuntu Distribution installed on WSL and accessed through Windows Terminal, version control with Git and GitHub, automation with GitHub Actions, static hosting with GitHub Pages, Python scripting to customize and enhance the functionality of GitHub Marketplace Actions, and Python's UNITTEST framework.

In the following sections, we provide a detailed description of the project and the concepts covered. We then reflect upon the successes and challenges we faced and report this experience so that other students at a similar level in their computer science education might find comparable opportunities to study these software engineering concepts earlier than they otherwise might while also attaining practical lab experience to supplement their traditional high school computer science curriculum.

## 2 RELATED WORK

Much has been written about the advantages of introducing version control with Git and GitHub in the classroom [4, 5, 6]. The concept of test-driven-learning, which relates to the software engineering concept of test-driven-development and advocates for demonstrating the use of automated tests alongside teaching programming concepts early in students' computer science education, is also found in the literature [7, 8]. Our experience corroborates the findings in the literature that an early introduction to version control and automated testing is advantageous but demonstrates a unique experience where the concepts are presented in a research lab setting that culminates with students submitting pull requests to deploy their automated tests to real open-source projects.

## 3 PROJECT PREREQUISITES

We will now take a closer look at each part of the project's development. Throughout the summer, our meetings took place using Zoom at intervals of one or two meetings per week and each session lasted between one and two hours.

### 3.1 Tools and Installation

After introductions, a discussion about scheduling, and a discussion about what we planned to accomplish over the summer, we decided upon and installed our development tools to ensure a consistent development environment. This involved stepping through the WSL installation process and setting up Ubuntu user accounts. We then installed Windows Terminal as our command line application and Visual Studio Code as our preferred text editor. We then visited github.com to explore GitHub's features and establish user accounts.

### 3.2 Linux Shell Familiarization

At our next meeting, we explored Windows Terminal and entered basic commands to navigate the directory structure. Basic commands such as *ls*, *pwd*, *cd*, *mkdir*, *touch*, *mv*, *cp*, and *clear* were discussed and demonstrated. We practiced creating directories, creating files, moving files, and displaying a directory's contents. In addition to navigating the command line on one's local machine, automation with GitHub Actions requires command line navigation on GitHub's runners. A familiarity with the command line will therefore be needed later in the project.

### 3.3 Introducing Version Control

For the next few meetings, we focused on learning version control. To assist with this learning and because the students had previous experience writing Python, we implemented a simple calculator application in Python. The main function provided a menu for users to choose arithmetic operations and to input operands. Each operation such as addition, subtraction, and raising a number to a power was a separate function that received the requisite operands as parameters and returned the result. A skeleton of the Python code was provided, with unimplemented function stubs that all returned zero.

The students learned to create a new repository on GitHub and clone the new repository to a directory on their local machine. Git was introduced using terminal commands rather than through a GUI to provide further command line practice, which will be useful when executing commands on GitHub's runners. Upon cloning the remote repository, we discussed the concepts of distributed versus centralized version control and discussed the pros and cons of each system. The students added the new calculator skeleton to the repository and pushed it to the remote. The Git commands *clone*, *pull*, *status*, *log*, *add*, *commit*, and *push* were introduced and applied.

Since two students were working on the project, one student implemented the addition function, and the other implemented the subtraction function. Each committed their local changes and pushed those changes to the remote. Each screen-shared while implementing the changes, and we noted how the second person to push needed to pull the changes already made to the remote before pushing one's own. It was also noted that having made changes to different functions, the differences could be reconciled automatically without giving rise to a merge conflict.

We next discussed the concept of branching and the benefits of developing features on specific feature branches. We created separate branches for the implementation of the multiplication and power functions. One student implemented *multiplication* and the other implemented *power*. Continuing to practice staging and committing files, we committed our local feature branches and learned to establish remote tracking for new branches. *git diff* was also introduced to compare un-staged changes with the most recent local commit and to compare other commits using the relevant commit hash. We then applied *git merge* to locally merge our feature back into our local main branch. Since each student had changed different parts of the code, it was again noted that the merge was automatic and gave rise to no merge conflict.

Next, we specifically edited the code to create and demonstrate a merge conflict. Having both edited the same section of the code, the students learned to manually resolve merge conflicts.

```python
# first student implementation
def divide(a, b):
    return a / b

# second student implementation
def divide(a, b):
    if b == 0:
        raise ValueError("divide by zero error")
    return a / b
```

**Figure 1: Students editing the same division function to purposefully give rise to a merge conflict.**

We concluded our introduction to version control with a look at pull requests and forks. In addition to merging branches locally, we discussed a typical workflow in which a developer without push access to the main branch would propose changes by submitting a pull request. We also looked at creating repository forks and saw how pull requests could be submitted across forks. The final calculator function, a square root function, was merged into the main branch through the submission and approval of a pull request.

To prepare for the next phase of our project, the students created their own forks of the JETSCAPE [1] and GOMC [2] application repositories as well as the JETSCAPE and GOMC website repositories. Most of the time spent writing these tests would pertain to the website repositories.

## 3.4 Introducing Continuous Integration

To introduce the concept of continuous integration, we created a simple Python script that uses the *requests.get* method to query a website and return a status code. The script can be run manually at the command line and accepts a URL as a command line argument. If the status code returned is 200, a success message is printed, and the program exits normally. If another code is returned or if there is no response from the server, the program terminates having raised an exception. We tested the program manually on popular websites like google.com and on URLs we knew to be invalid.

We then discussed the usefulness of having a test run automatically either on a set schedule or whenever there are changes to a specified branch. Since there are two students and two websites (JETSCAPE [1] and GOMC [2]), each student focused on a specific website and worked on their locally cloned fork. We committed the Python script that checks for reachability and discussed the YAML file steps required to generate the automation. We decided that the test should run on push operations and pull requests to the main branch and run once per day at a specific time. We discussed the job steps including running on *ubuntu-latest*, checking out the current repository using GitHub's checkout action, and executing terminal commands on GitHub's runners to install the *requests* library using *pip*. We discussed that this step would be performed every time the workflow executes because each job runs on a new instance of *ubuntu-latest*. We then added the job steps to navigate to the *tests/* folder on the runner and to execute the *python3* command calling the URL reachability script. The students committed, pushed to the remote forks, and checked the Actions tabs on GitHub to review the running tests and their respective logs.

Equipped with prior knowledge of the Python language and having now gained a familiarity with the command line, version control, and automating workflows with GitHub Actions, the students were ready to start researching and experimenting with writing and deploying the automated tests.

# 4 PROJECT IMPLEMENTATION

To meet the project's goal of creating an experience that emphasizes learning while also providing students an opportunity to write and deploy meaningful tests, we chose the following three tests to implement: an HTML validator, a link checker, and a series of unit tests to ensure the correctness of the JETSCAPE [1] XML reader, which is a function of the JETSCAPE framework that parses input parameters. Since the first two tests don't involve unit tests, an introduction to unit testing and the Python UNITTEST framework can be postponed until after the first two tasks are completed.

## 4.1 HTML Validator

The JETSCAPE [1] and GOMC [2] websites are both static sites intended to convey information about the JETSCAPE and GOMC scientific computing applications. The websites are written with HTML, CSS, and JavaScript, and the sites are hosted and deployed using GitHub Pages. The sites can be updated through pull requests to the repositories' respective main branches, and members of the collaboration can propose pull requests. Since the sites are collaboratively maintained, it would be useful to have an automated test to ensure that all HTML code conforms to the HTML5 standard. Invalid HTML can often result in small easy-to-miss cosmetic inconsistencies, so this automated test would prove useful.

The students explored the GitHub Actions Marketplace and together we read through the documentation of a Marketplace Action [10] that looked relevant to our task. The Marketplace Action referenced and extended an open-source HTML5 Validator [11] built with Python and available through *pip*. We installed the validator locally to test it on our repository code, and purposefully added HTML tag typos to force the tests to fail.

After thoroughly testing the validator locally, we then collaboratively wrote the YAML file to automate running the test on push operations and pull requests. Having only previously seen the example YAML file to check whether a website is reachable, we referred to that example file as a starting point, removed what wasn't relevant for this workflow and added the step to call the validator Marketplace Action [10].

The students then pushed their committed changes to their respective forks and viewed the running tests in the Actions tab on GitHub. An issue arose regarding how to format the path to the repository checked out on GitHub's runner. Since the path to the repository on the runner is different from the path to the repository on one's local machine, the students gained experience reading the test logs and examining why the test that had passed locally was failing on GitHub's runners.

```yaml
jobs:
  validate:
    runs-on: ubuntu-latest
    steps:

      - name: Checkout Repository
        uses: actions/checkout@v2

      - name: HTML5 Validator
        uses: Cyb3r-Jak3/html5validator-action@v7.2.0
        with:
          root: ./
          css: true
```

**Figure 2: A partial implementation of the YAML file automating the HTML validator.**

Although the HTML validator was a simple test to automate, we consider it to have been a good starting point. The Marketplace Action [10] does most of the work and successfully validates the repository's code without requiring any additional scripting, so the students could focus on testing the implementation and writing the YAML file to handle the automation. And while Git operations remain a new concept, the students continued to develop fluency with each subsequent sequence of commits and pushes.

## 4.2 Link Checker

The next test built upon the skill acquired implementing the HTML validator and required some additional scripting and troubleshooting.

A link checker is a useful website debugging tool that identifies broken links. We discussed the usefulness of setting the link checker to run on a schedule in addition to on pushes and pull requests. While the link checker will catch broken links caused by a developer mistyping a URL in the code, it is also reasonable to expect that a presently working link could at some point stop working. Many website links don't just point to internally hosted pages and content but point to external conference pages and journal articles. Links to old conference sites can change or simply go offline. To regularly check the validity of every link would be a tedious task to perform manually. With respect to the JETSCAPE [1] and GOMC [2] sites, no one was checking links regularly and once implemented, our tool identified several broken links on each site.

The students began by researching GitHub Marketplace Actions and other available open-source tools as they did for the HTML validator. We found a Python tool, *LinkChecker* [12], that supported a command line interface. We installed the tool locally and tested it. The tool takes a URL as a command line argument, recursively visits pages, and checks internal links reachable from the initial URL. A flag can be set to instruct the tool to also check external links.

In our testing, we discovered that when passing our websites' root domain, *https://jetscape.org* or *https://gomc-wsu.org* respectively, the links on the *index.html* page were checked, but no other site page was visited. The students investigated and realized that the navigation bars were inserted with JavaScript from text on a separate *nav.html* page, and the tool itself does not render JavaScript. To solve the issue, we passed the URL of the *nav.html* page instead of the domain's root address. The students examined the test logs and determined that every *.html* page was now being visited, but many of the external conference and journal links specifically on the JETSCAPE website were not being checked.

JETSCAPE's site design uses several JSON files where the conference and publication data are stored. JavaScript parses the JSON data, which includes the conference and publication links, and writes them to the relevant tables on page load. Therefore, the tool passes over the most important links we would want to check. To solve the problem, we decided to supplement the open-source tool, which successfully checks the links specifically written in the HTML files, with our own link checking script designed to identify and check links found in the JSON files.

A Python script skeleton was provided to the students with function stubs to be implemented. Providing a code skeleton was familiar and consistent with the methodology used to introduce version control with the calculator application. Code from the URL reachability script and YAML file that introduced GitHub Actions was also applicable and could be referenced when implementing the link checker. The students were guided to produce a Python script that accepted a directory path as input and parsed the text of every JSON file found in that path to identify and check URLs. We successfully tested this script locally, passing the path to our repository's data folder where the relevant JSON files were stored.

We then wrote our YAML file to automate a test that used the original open-source link checker to test the links written explicitly in the HTML files as well as our own Python script to check the links identified in the JSON files. We then tested the automation by including purposefully broken links. During this testing, we realized that the original open-source link checker was checking the deployed site rather than the code we had just changed and were about to deploy. This was not what we intended. We wanted to check the code we were pushing to ensure that proposed changes didn't introduce new problems. To solve this issue, we amended our YAML file to launch a local server on the GitHub runners. Instead of passing the deployed site address to the link checker, we passed the address of the site launched on the GitHub Actions runner's local server. This solution required a discussion and demonstration of how to test websites using a local server and how to run programs as background processes at the command line. Completing these tasks, the students now had multiple experiences practicing Git operations, working at their local command prompts, and writing commands as YAML file job steps to facilitate automation with GitHub Actions.

```yaml
- name: Checkout Repository
  uses: actions/checkout@v2

- name: Check JSON File Links
  run: |
      cd $GITHUB_WORKSPACE/tests
      python3 linkcheck.py $GITHUB_WORKSPACE/data

- name: Run Link Checker
  run: |
      cd $GITHUB_WORKSPACE/
      nohup python3 -m http.server </dev/null >/dev/null 2>&1 &
      export LOCAL_HOST=$(hostname -i)
      echo "running link checker..."
      linkchecker --verbose --check-extern http://${LOCAL_HOST}:8000/nav.html
```

**Figure 3: A partial implementation of the YAML file automating the link checker.**

## 4.3 Introducing Unit Testing

The third automated test required an introduction to the concept of unit testing. For this introduction, we returned to the calculator application used to introduce version control. Building upon that code, we explored Python's UNITTEST framework and created a test class with functions to test the calculator's arithmetic operations. It was noted how several test cases could be executed with one simple call to a Python script. We then wrote a YAML file to automate running those tests on pushes and pull requests to the calculator repository.

## 4.4 Unit Testing the XML Reader

Our final automated test applies to the JETSCAPE [1] code repository instead of the website repositories. The JETSCAPE application provides an XML reader to receive a user's input parameters. JETSCAPE maintains a *main.xml* file in which default parameters are set. A user can provide a separate *user.xml* file to override some or all the *main.xml* tags. To prevent typos that could arise in the *user.xml* files, the JETSCAPE application will exit with an error message if the *user.xml* file contains a tag that is not also included in the *main.xml* file.

Many example *user.xml* files are included in the repository to demonstrate different use cases, and as new features are added, developers can include new example XML files. The automated test described here will use Python's UNITTEST framework to call JETSCAPE's XML reader and test every example *user.xml* file included in the repository.

The requirements of this test exhibit comparable functionality to the link checker. While the link checker needed to identify every JSON file in a directory, here we look for every *user.xml* file in a subtree of directories. This offers students an opportunity to revisit and review logic and syntax seen for the first time with the link checker test.

# 5 INSIGHTS AND REFLECTIONS

In this section, we reflect upon what we learned, what worked well for us, what challenges we faced, and how a project such as this could be adapted and scaled in the future. We share our outlook that exposure to version control and automated software testing early in a student's computer science education can provide a strengthened ability to work through errors, find and resolve bugs, and write cleaner code.

## 5.1 Adaption, Future Work, and Scalability

This project and the automated tests we developed applied specifically to the scientific computing applications developed in connection with our lab. We wanted to give our volunteer students an opportunity to gain practical experience and meaningfully contribute to these scientific computing applications while also being introduced to new concepts and learning new skills. The project websites provided a good place to start because the students could immediately begin learning the relevant software engineering concepts without requiring prerequisite domain science knowledge in physics (JETSCAPE [1]) or chemical engineering (GOMC [2]).

We see value in generalizing this content and scaling it to benefit other students for other applications. The repository management and automated testing concepts can apply broadly to any informational websites from local small businesses to a student's own promotional website showcasing work to prospective employers.

A challenge we faced with respect to developing this project is that these scientific computing applications and other research projects active in our lab would require prerequisite domain science knowledge, knowledge of algorithms and data structures, or knowledge of parallel programming and high-performance computing. A project to develop these automated tests was something we could achieve without overwhelming background requirements. The XML reader was chosen to unit test because it was a part of the application that could be understood without domain science knowledge. However, implementing unit tests to validate example input files is not, in our view, the most pedagogical use case for unit testing. For a future endeavor, it may be more appropriate to introduce unit testing to validate the correctness of a function itself rather than the validity of an input file.

We faced an additional challenge in that, in our view, we were unable to provide an overview of the project at the outset. Much of this project's development was experimental and we weren't sure how long it would take to complete each task. We also didn't have upfront knowledge about which of the available utilities in the GitHub Actions Marketplace would work for us; therefore, the project involved researching the available tools and encountering trial and error as we progressed. While this provided exposure to the experimental environment of a laboratory, the students might have felt better prepared if a more complete overview of the project's scope, schedule, and tools were given at the outset. An introductory handout of commands and concepts could have provided the students with a more actionable roadmap, study guide, and measure of their progress throughout the project.

## 5.2 Advantages of Early Learning

Software engineering concepts such as version control, unit testing, and continuous integration often aren't introduced until several semesters into an undergraduate student's studies and after multiple programming courses. Having worked with many students at the undergraduate level, it is our observation that students often resort to saving multiple versions of their code under different file names, creating a disorganized working environment that could be improved with a brief introduction to version control. Practicing version control early in one's programming studies could lead students to developing fluency in this important skill well before graduation or internships [9]. A knowledge of software testing – especially unit testing – could lead beginning programmers to think more cleanly about program design, and a requirement to write unit tests could lead beginning students to write more succinct and modular functions as opposed to the lengthy and sprawling functions often observed in beginning students' code.

## 6 CONCLUSION

For motivated independent study students, learning software engineering concepts such as version control and continuous integration will benefit them in their future studies. As demonstrated here, this content is accessible to students even at an early point in their computer science education. A special honors project or a supplemental one-credit-hour tools and techniques course taken as a corequisite to the traditional introduction to programming track could be a next step to further develop, scale, and generalize the material and unique format devised and embarked upon for this project.

## ACKNOWLEDGMENTS

We would like to acknowledge Loren Schwiebert at Wayne State University for his assistance, feedback, and advice throughout the development and implementation of this project. We would also like to acknowledge the JETSCAPE Collaboration [1] and GOMC [2] for providing the source repositories and applications on which our automated tests were developed. We also acknowledge *LinkChecker* [12], *GitHub Actions HTML5 Validator* [10], and *svenkreiss/html5validator* [11] as the marketplace and open-source utilities discussed and applied as part of this project.